\documentclass[sigchi]{acmart}

\copyrightyear{2019}
\acmYear{2019} 
\setcopyright{iw3c2w3}
\acmConference[WWW '19]{Proceedings of the 2019 World Wide Web Conference}{May 13--17, 2019}{San Francisco, CA, USA}
\acmBooktitle{Proceedings of the 2019 World Wide Web Conference (WWW '19), May 13--17, 2019, San Francisco, CA, USA}
\acmPrice{}
\acmDOI{10.1145/3308558.3313638}
\acmISBN{978-1-4503-6674-8/19/05}

\usepackage{url}
\usepackage{latexsym}
\usepackage{amsmath}
\usepackage{graphicx}
\usepackage{subcaption}
\usepackage{booktabs}
\usepackage{color}

\newcommand{\xhdr}[1]{{\noindent\bfseries #1.}}

\newcommand{\cut}[1]{}

\newcommand{\sectionrule}{\addlinespace[1ex]}

\title{Trajectories of Blocked Community Members: Redemption, Recidivism and Departure}
\renewcommand{\shorttitle}{Trajectories of Blocked Community Members}

\author{Jonathan P. Chang}
\affiliation{%
  \institution{Cornell University}
  \postcode{14850}
}
\email{jpc362@cornell.edu}

\author{Cristian Danescu-Niculescu-Mizil}
\affiliation{%
  \institution{Cornell University}
  \postcode{14850}
}
\email{cristian@cs.cornell.edu}

\begin{document}
\begin{abstract}

Community norm violations can impair constructive communication and collaboration online.
As a defense mechanism, community moderators often address such transgressions by temporarily blocking the perpetrator.  
Such actions, however, come with the cost of potentially alienating community members.  
Given this tradeoff, it is essential to understand to what extent, and in which situations, this common moderation practice is effective in reinforcing community rules.

In this work, we introduce a computational framework for studying the future behavior of blocked users on Wikipedia. 
After their block expires, they can take several distinct paths: they can reform and adhere to the rules, but they can also recidivate, or straight-out abandon the community.  
We reveal that these trajectories are tied to factors rooted both in the characteristics of the blocked individual and in whether they perceived the block to be fair and justified.
Based on these insights, we formulate a series of prediction tasks aiming to determine which of these paths a user is likely to take after being blocked for their first offense, and demonstrate the feasibility of these new tasks.
Overall, this work builds towards a more nuanced approach to moderation by highlighting the tradeoffs that are in play.

\end{abstract}

\maketitle

\section{Introduction}
\label{sec:intro}

The health of online communities is often threatened by individuals that violate their norms and rules \cite{kiesler_regulating_2012}.
Communities have limited defense mechanisms against such offenses, ranging from the  deletion of problematic comments \cite{chandrasekharan_internets_2018}, to temporarily blocking the perpetrators, to outright excluding them \cite{jhaver_online_2018}.
Such actions, however, come at the cost of potentially alienating valuable community members and raise issues of freedom of speech, fairness and bias \cite{king_how_2013,shen_perceptions_2018,jhaver_online_2018}. 
Considering their potential negative impacts, it is important to understand to what extent, and in which situations, these moderation mechanisms are effective at enforcing community norms.

In this work we take a first step towards answering these questions by focusing on \textit{temporary blocks} in the Wikipedia community of editors.  
Unlike other moderation actions, these have the explicitly stated purpose of ``deterring any future possible repetitions of inappropriate conduct'' by the perpetrators, while reintegrating them in the community (thus the temporary nature of the block).  
However, following their initial block, nearly half of first-time offenders either recidivate---18\% of established members violate the rules again within 6 months of the block---or straight-out abandon the community---30\% of established members depart within 6 months (a third of which depart during the block). 
This discrepancy between the stated goal and the empirical outcomes of temporary blocks motivates an investigation into the factors that determine the likely trajectory a community member will take after their block expires: redemption, recidivism or departure. 

We ground our investigation in offline theories of deterrence and defiance \cite{sherman_defiance_1993}, which point to at least three broad classes of factors determining future compliance: those pertaining to the characteristics of the perpetrator \cite{tonry_learning_2008}, those 
pertaining to the actual severity of the punishment \cite{grasmick_deterrent_1980, klepper_tax_1989} and those pertaining to whether the perpetrator perceives their treatment as fair \cite{paternoster_fair_1997}.  
While relying on these theories to guide our selection of factors and potential confounds to consider in our analysis, we refrain from drawing any direct parallels between offline and online rule-enforcement strategies given their distinct nature and goals. 

\xhdr{Individual characteristics}  
Behavior of community members and their adherence to community norms has been shown to vary according to their level of involvement with the community 
\cite{danescu-niculescu-mizil_no_2013,cheng_antisocial_2017,ribeiro_characterizing_2018, halfaker_dont_2011}.
We find that individuals that exhibit a high level of community involvement (e.g., have interacted with more users over a longer period of time) are less likely to abandon the community during a temporary block. 
On the other hand, these individuals are also more likely to recidivate soon after the block.  
This suggests that users with high community involvement are more immune to the effects of blocks, both undesired (departure) and desired (reforming). 

\xhdr{Block duration}  Existing literature on offline rule compliance varies in their accounts of how the nature of the punishment determines the likelihood of recidivism \cite{tyler_why_1990}.
The traditional (institutional) perspective is that the severity of the punishment is the main factor determining compliance~\cite{grasmick_deterrent_1980, klepper_tax_1989}.
A contrasting (normative) perspective, however, argues that perhaps a more important factor is whether the perpetrator regards the punishment as being fair and justified \cite{makkai_reintegrative_1994, tonry_learning_2008}.  
In the context of the Wikipedia community, we find that even though longer---and thus more severe---blocks come at the cost of higher departure rates, they do not appear to have any effect on recidivism.   

\xhdr{Perceived fairness}
To capture whether an individual perceives their initial block as fair, we leverage a mechanism specific to online moderation: block appeals.
Blocked individuals can contest a block by opening a discussion with Wikipedia moderators during the block.  
In turn, moderators have the option to lift the block early---i.e., to \emph{unblock} the individual---as a result of these discussions.
This process provides us with a glimpse into a blocked individual's perception of fairness: we can analyze their language as well as the outcome of the appeal. 

Using this methodology, we find that users who perceive the block to be unfair have an increased rate of recidivism, while those that acknowledge their wrongdoing and apologize have a decreased rate.  Furthermore, even though the block duration itself does not appear to affect recidivism, the post-hoc reduction of this duration 
by a moderator 
drastically diminishes recidivism rates, perhaps by improving the user's perception of the moderation system as being fair and reasonable. Combined, these results suggest that the effectiveness of temporary blocks is not so much dependent on their actual severity, but rather on the blocked user's perception of how appropriate the block is.

To summarize, in this work we:

\begin{itemize}
	\item investigate the effectiveness of temporary blocks by considering possible trajectories an individual can take after their block expires;
	\item reveal factors that are indicative of these future trajectories, reflecting both the 
	engagement patterns
	 of the blocked individual, as well their perception of the fairness of the block;
	\item analyze the relative importance of each of these factors by using them in a series of forecasting tasks aiming to determine which path a user is likely to take after their first block.
\end{itemize}

More broadly, we propose a computational framework for analyzing the outcomes of a common moderation practice in online communities.  
In doing so, we expose the underlying tradeoffs and provide the means to  
further explore factors that mediate them.

\section{Blocks on Wikipedia}
\label{sec:data}

While Wikipedia is primarily known as an online encyclopedia, it also plays host to a vibrant community of editors who continually write new articles and improve existing ones.
To support this community, Wikipedia has a feature known as \emph{talk pages}: special pages on which editors can discuss a particular article or Wikipedia policy, or simply unwind with casual conversation.
Every Wikipedia article has an associated talk page, on which editors can discuss proposed edits to the article.
Similarly, every registered user has an associated \emph{user talk page}, where other editors can engage directly with the respective user on issues pertaining to their general behavior.  This is where the social functions of the community surface: editors discuss inappropriate behavior and norm violations, ask for support, offer advice or encourage each other.
In this work we use the complete conversational history between English Wikipedia editors on both article and user talk pages.  With over 90 million conversations between 4 million users on 24 million talk pages, this is one of the largest collections of public conversations \cite{hua_wikiconv_2018}.

Like many online communities, Wikipedia has a moderation system aimed at imposing community rules.  What makes Wikipedia particularly suitable for our proposed investigation is that its rules apply uniformly to the entire community and moderation is entirely transparent to the public \cite{forte_scaling_2008}.  This is unlike other large communities, such as Reddit, where rules can vary drastically between sub-communities and where aspects of the moderation remain private \cite{chandrasekharan_internets_2018}.

\xhdr{Blocking mechanism} Wikipedia moderators 
(formally referred to as administrators, though we will continue to use the term ``moderator'' as we are focused on their role in maintaining community norms) %
are elected from among community members through public elections \cite{burke_mopping_2008,leskovec_governance_2010}.  
Among other privileges, moderators have the ability to block users from making any edits or comments.  
They can either block a user indefinitely, in cases of accounts that are clearly used for bad-faith purposes, or temporarily if they see the user as a potentially valuable member of the community that has momentarily gone awry. 
In fact, the stated purpose of temporary blocks is ``to deter any future possible repetitions of inappropriate conduct'', thus keeping the door open for reintegration after the expirations of the blocks.\footnote{\url{https://en.wikipedia.org/wiki/Wikipedia:Blocking_policy}} 
Our focus in this work is on temporary blocks (henceforth simply called blocks), and their effectiveness in reforming blocked users.

When temporarily blocking an individual, moderators are required to notify the user by posting a message on their user talk page. 
This notification includes the reason for the block and its duration. The duration of the block is at the discretion of the moderator, and can vary greatly---the standard deviation of block durations is 194 days.
However, this high variance is largely the result of a small set of extreme outliers, and in practice most blocks are very brief---the median duration is 1 day. 

For the duration of the block, the targeted user may not contribute to Wikipedia articles or make any comments on talk pages, with the sole exception of their own user talk page, where they are allowed to discuss the block.  

\xhdr{Block reason} 
When moderators block a user, they must cite a reason that reflects one of the community rules.
Wikipedia groups these reasons under two broad categories: \emph{protection} and \emph{disruption}. 
The former concerns behaviors that may put Wikipedia editors or Wikipedia itself at risk, such as making legal threats, releasing personal information or violating copyright law.  
The latter concerns the health of the community---or, as Wikipedia phrases it, maintaining a ``civil, collegial atmosphere''.  
We identify a subset of four such disruption block types relating to breaches of community norms (rather than to legal concerns specific to Wikipedia): personal attacks and incivility, harassment, edit warring, and disruptive editing.  
These reflect norms that can be broken even by established and valuable members of the community, unlike spam or vandalism which most often come from illegitimate, disruption-only accounts. %

\begin{table}
    \centering
    \begin{tabular}{lr}
        \hline
        Total number of blocks & 104,245 \\
        \sectionrule
        Number of blocked users & 72,332 \\
        Number of blocking moderators & 1,706 \\
        \sectionrule
        Users with block reason in disruption subset & 21,043 \\
        Users with first block in disruption subset & 18,909 \\
        \sectionrule
        Users remaining after minimum activity filters & 6,026 \\
        \hline
    \end{tabular}
        \caption{Statistics of the blocked users dataset.}
    \label{tab:table_dataset_stats}
\end{table}

\xhdr{Blocked users dataset} We extract information about the nature of the blocks from the \emph{block log}, a centralized and public record of all blocks that have ever happened on Wikipedia containing over 136,000 actions (including both blocks and manual block modifications) at time of collection. We automatically extract from this log the block's duration,\footnote{We merge consecutive blocks and account for post-hoc reduction in block duration.} its type, its date,\footnote{In our analysis we discard blocks happening in the first five years of the community, allowing for  Wikipedia community norms to stabilize~\cite{halfaker_rise_2013,forte_scaling_2008}.} and the involved users 
(blocking moderator and blocked individual).
We seek to focus our analysis on actual members of the community as opposed to disruption-only accounts, because such accounts are made in bad faith and hence have no reason to reform once caught \cite{geiger_work_2010}. %
To this end, we discard accounts whose first block was not one of the four disruption block types previously described, and impose a minimum activity before the first block: at least one month and eight comments.
Finally, we join the resulting block meta-data with the conversational record. Details of the resulting dataset are summarized in Table \ref{tab:table_dataset_stats}. We make this processed dataset publicly available\footnote{As part of ConvoKit (\href{http://convokit.cornell.edu}{convokit.cornell.edu})} to encourage the study of moderation in a community where all activity is by design fully transparent and public.\footnote{All editors are expressly informed, during all edits, that ``Work submitted to Wikipedia can be edited, used, and redistributed---by anyone'', where ``work'' is defined as content on any Wikipedia page, talk pages included. See \url{https://en.wikipedia.org/wiki/Wikipedia:Ownership_of_content}.}

\section{Trajectories after the block}
\label{sec:trajectories}
After their block expires, a user can take several distinct trajectories: they can reform and adhere to the rules, but they can also recidivate, or straight-out abandon the community.  
Here we formally define these (un)desired outcomes and reveal their relative prevalence.

Throughout, we focus on the very \emph{first} block a user receives, thus discarding any potential lingering effects of previous blocks and putting all users in our dataset on equal ground at the time of investigation.
We consider a maximum future horizon of 6 months, and thus are constrained to discard users whose initial block takes place less than 6 months before the end of our dataset (June 2018) since for these we cannot know what the future holds.  
This is a common solution for limited horizon in longitudinal studies 
\cite{danescu-niculescu-mizil_no_2013}. 

\xhdr{Departure} Wikipedia and other collaborative communities have traditionally struggled to retain their members \cite{halfaker_rise_2013}.  
Thus, the risk of alienating users needs to be taken into consideration when evaluating the efficacy of moderation.
This is particularly true for temporary blocks, which by their very nature aim at re-integrating the targeted user into the community rather than discouraging them to contribute again (for which a moderator would use a permanent block).  

We consider a user to \emph{depart} from the community at the time of their last comment on community talk pages; we use this instead of their last edit on an article as the indicator of quitting because we are specifically interested in their engagement with the community. 
We impose the same 6 month maximum future horizon as we did when considering recidivism.
Out of all 6,026 %
blocked users selected for this study (Section \ref{sec:data}), 
30\% depart in the 6 months following their first block, and 10\% depart \emph{during} the block itself.

\begin{figure}
    \includegraphics[width=.4\textwidth]{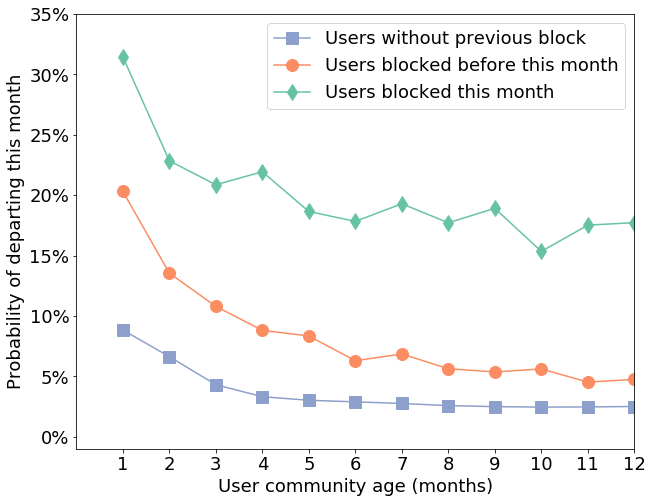}
    \caption{Probability of a user quitting at a given community age (in months), given that they were never blocked before (blue squares), blocked anytime prior to that month (orange circles) or blocked in that same month (green diamonds). The trends continue beyond the 12 months illustrated here.}
    \label{fig:figure_quit_trajectories}
\end{figure}

To illustrate the connection between departure rates and the block itself, Figure \ref{fig:figure_quit_trajectories} compares the probability of a user abandoning the community at different stages in their community life (i.e., time since their first activity), conditioning on whether (and when) they were blocked before.   
We first note that departure is not independent from blocking events (compare blue squares with orange circles): users are much more likely to depart if they were blocked before, especially in the first months of their life in the community.  
But even controlling for the fact that a user was previously blocked (compare orange circles with green diamonds), they are much more likely to depart during the month of their block.  This strong recency effect shows that blocking and departure are events that are also temporally connected. %

\xhdr{Recidivism} A previously-blocked user \emph{recidivates} if they are blocked again for any other breach of community rules, and \emph{reforms} if they never get blocked again.
There is a wide variance in how long it takes recidivst users to re-offend.
While some get blocked again within a week, others remain ``clean'' for years before re-offending.
Only 47.7\% of blocked users in our dataset have no (blocked) offenses after their first block.
Among the users who do have a second known offense, the median time to reoffense is 32 days.

To study recidivism at different temporal scales, we consider two categories of users: \emph{long-term recidivists} are those who re-offend within the next 6 months after their initial block, and \emph{short-term recidivists} are the subset who re-offend within the first week after their first offense.  
Out of all blocked users selected for this study, 38.5\% breach the norms again in the 6 months following their initial block; and 15.4\% recidivate in just one week after the initial block.

\begin{figure}
    \includegraphics[width=.4\textwidth]{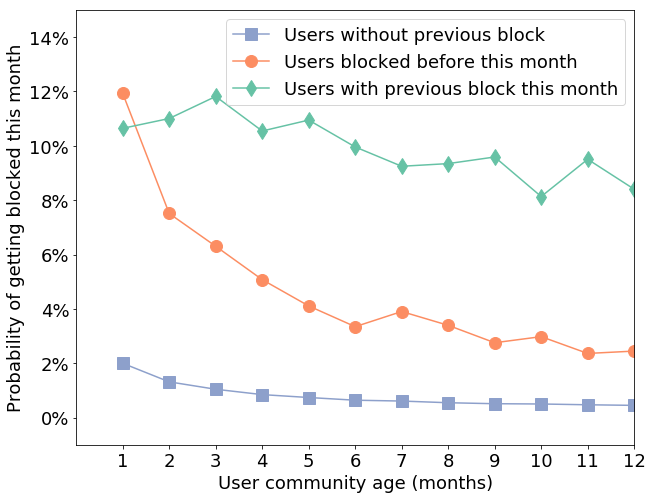}
    \caption{Probability of a user getting blocked at a given community age (in months), given that they were never blocked before (blue squares), previously blocked anytime prior to that month (orange circles) or previously blocked in that same month (green diamonds).} 
    \label{fig:figure_block_trajectories}
\end{figure}

To put recidivism rates into perspective and differentiate the phenomenon from general rule violations, Figure \ref{fig:figure_block_trajectories} compares the probability of a user getting (re)blocked at different stages in their community life, conditioned on whether (and when) they were already blocked or not.  
Perhaps unsurprisingly, individual block events are not independent: the probability of being blocked is much higher for individuals who were already blocked before than for those that were not, even after controlling for community age (compare blue squares with orange circles). 
We also observe a dramatic decrease in probability of recidivism during the first year of a user's community age.
Importantly, this effect is specific to recidivism, rather than just reflecting a tendency to get blocked in general---the probability of getting blocked for the first time (blue squares) stays relatively flat over time.
Finally, as in the case of departure, we also observe a recency effect (compare green diamonds with orange circles), showing that consecutive blocks are temporally related events. %

Having established a relation between blocks and departure or recidivism, we now move towards a more thorough analysis of the factors that are in play.  We will explore two broad classes of factors inspired by theories of deterrence and defiance \cite{sherman_defiance_1993}.  The first class of factors concerns the characteristics of the perpetrator (Section \ref{sec:activity}): some types of individuals might simply be more likely to (re)offend or depart.  The second class of factors concerns the actual context of the block: its actual severity and how the perpetrator perceives it (Section \ref{sec:context}).

\section{User characteristics}
\label{sec:activity}

\begin{figure*}
    \centering
    \begin{subfigure}[t]{.4\textwidth}
            \includegraphics[width=\textwidth]{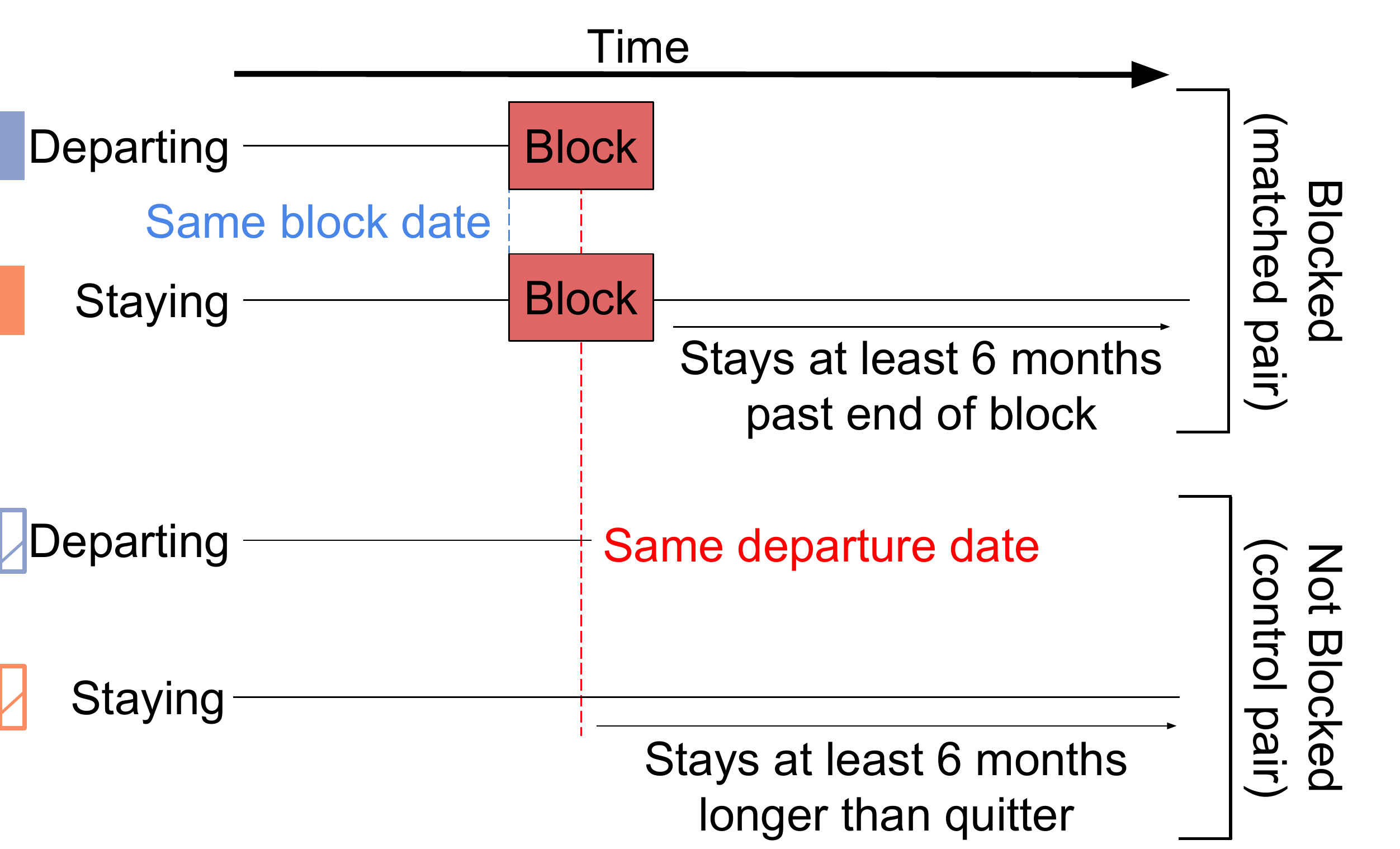}
        \subcaption{Departure matching and control}
        \label{fig:pairing_quit}
    \end{subfigure}
    \hspace{0.6in}
    \begin{subfigure}[t]{.4\textwidth}
               \includegraphics[width=\textwidth]{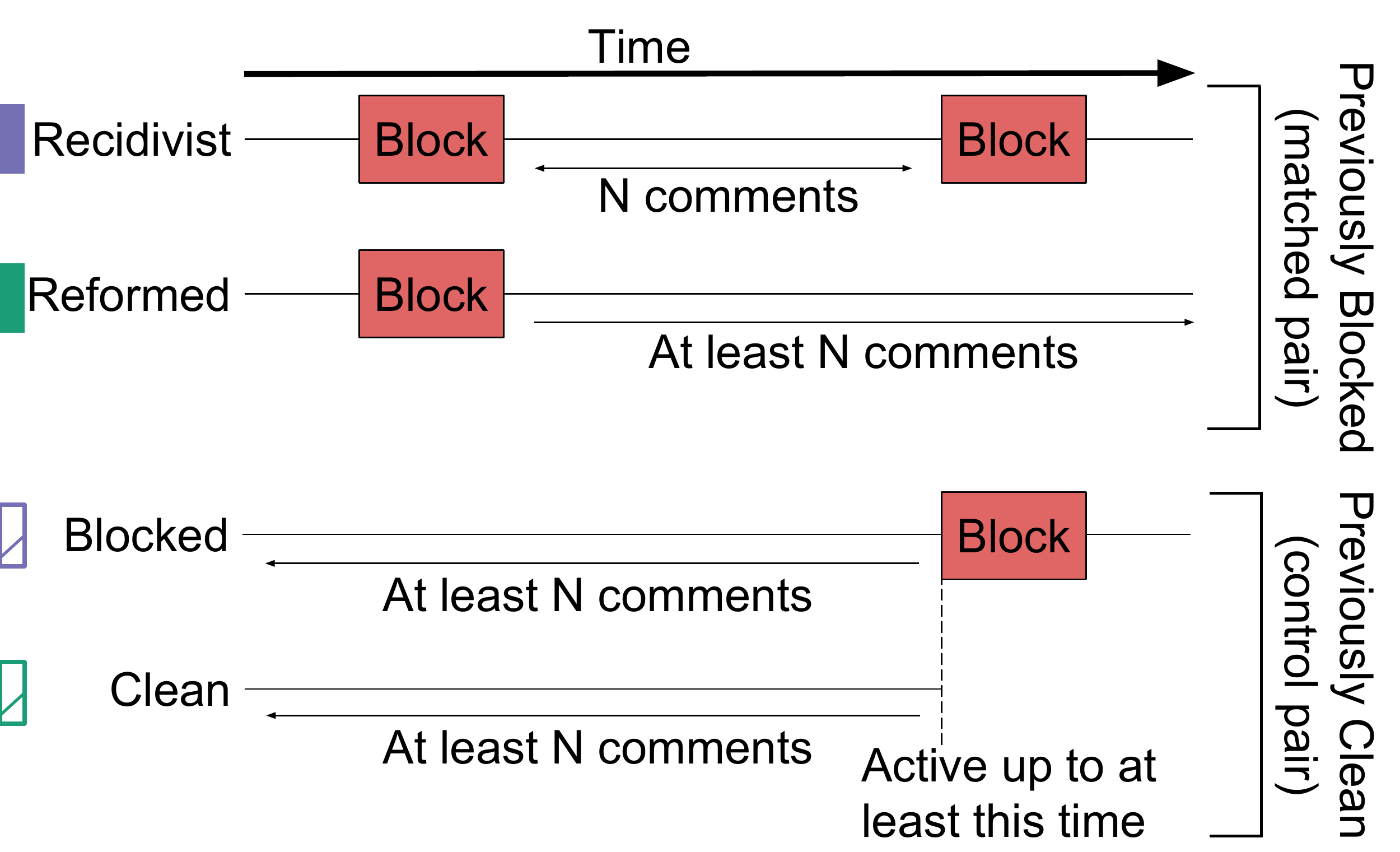}
        \subcaption{Recidivism matching and control}
        \label{fig:pairing_recidivism}
    \end{subfigure}
    \caption{We matched comparable blocked users with opposite outcomes: (a) departs or not during the first block; (b) offends or not after the first block. For each matched pair, we also select a control pair of users (with the same opposite outcomes) who were not blocked.}
    \label{fig:pairing}
\end{figure*}
We have already seen that community age is an important factor mediating the likelihood of departure and (re)offense: overall, newcomers are more likely than experienced users to follow one of these undesired paths after being temporarily blocked.  
Here we explore how these outcomes relate to other user characteristics, focusing on factors indicating a member's level and type of engagement in the community. 

We consider two broad types of features: how active the user is and how broad their social connections are in the community.  
To foreshadow our findings, we find that individuals that are more active in the community are overall more immune to blocks: they are both less likely to depart during a block, but also less likely to reform.  
However, the way an individual spreads their activity is also informative: if their activity is directed at a broader set of users they are more likely to either depart or reform.

\xhdr{Activity level}  We consider both the number of comments a user contributes to other users' talkpages (contributed activity), as well as the number of comments they receive from other users on their own talkpage (received activity) before the block.  To disentangle a user's activity from their community age, we normalize these measures by the number of days between the user's first recorded activity in the community and the time of the block.

\xhdr{Activity spread} An active individual can either engage a broad audience, or instead concentrate on a narrow set of individuals.  We capture the \emph{received activity spread} of an individual by counting the number of unique users who wrote comments on their talk page before the block, then normalizing it by the total number of such comments. %
A higher value of this ratio indicates brief interactions with a relatively large number of peers, while conversely, a smaller ratio indicates extended interactions with a relatively small group of peers.
Analogously, we define \emph{contributed activity spread} as the ratio between the number of other users' talk pages to which an individual contributed comments and the number of such comments they contributed.

\begin{figure*}
    \centering
    \begin{subfigure}[t]{0.225\textwidth}
        \includegraphics[width=\textwidth]{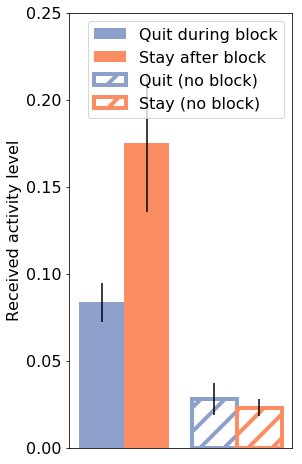}
    \end{subfigure}
    \begin{subfigure}[t]{0.225\textwidth}
        \includegraphics[width=\textwidth]{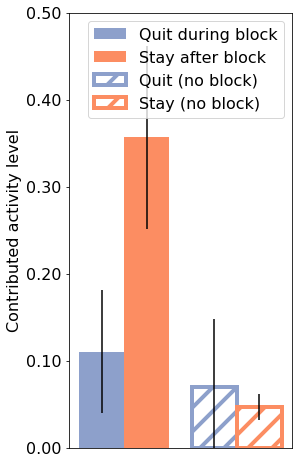}
    \end{subfigure}
    \begin{subfigure}[t]{0.225\textwidth}
        \includegraphics[width=\textwidth]{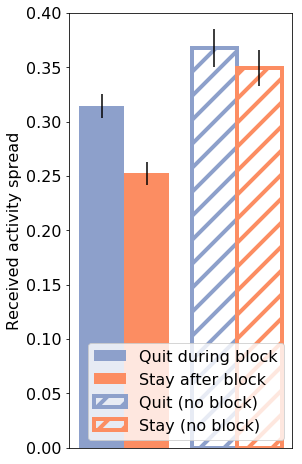}
    \end{subfigure}
    \begin{subfigure}[t]{0.225\textwidth}
        \includegraphics[width=\textwidth]{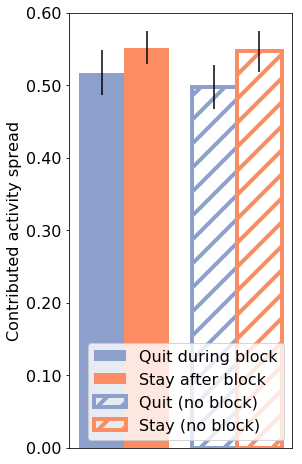}
    \end{subfigure}
    \caption{Difference between the community engagement patterns of departing and staying users. From left to right: receiving activity level,  contributed activity level, receiving activity spread, contributed activity spread. Solid bars compare blocked departing and staying users; hashed bars compare non-blocked departing and staying users. Throughout errorbars indicate 95\% binomial proportion confidence intervals.}
    \label{fig:figure_quit_social_features_comparison}
\end{figure*}

\xhdr{Departure}  To make meaningful comparisons between users who depart \emph{during} their first block and those who don't, hence eliminating trivial confounds, we use a matching technique inspired by causal inference methods \cite{rosenbaum_design_2010}.  
As illustrated in Figure \ref{fig:pairing_quit} (top), we pair each departing user with a user whose first block took place on the same date (within 1\% tolerance) but who stayed active for at least 6 more months (our maximum future horizon).  This procedure results in 1,156 paired departing and staying users.

As a secondary question, we also want to understand whether any differences we observe are specific to blocked individuals, or generally characterize community departure.  
To this end, for each matched pair of blocked departing-staying individuals, we construct a control pair of departing-staying individuals that were not blocked. 
This control pair consists of a non-blocked departing individual with the same departure date (within tolerance) as the blocked departing individual, and a staying individual who remained active at least 6 months past that date.
This process is also illustrated in Figure \ref{fig:pairing_quit} (bottom).

Figure \ref{fig:figure_quit_social_features_comparison} summarizes how our engagement factors relate to departure during the block (solid bars).  Departed users receive and contribute less daily activity in their life before the block than non-departed users. The activity they do receive comes from a broader set of users, potentially signaling they are not integrated into a tight social group.\footnote{These trends continue to hold, and remain significant, if we consider only the top half of active users in our comparison, thus separating them from the previously described trends in raw activity counts.} %
This is however not true for the spread of the activity they contribute, which is similar between departing and staying users.

We note that these relations between activity and departure are specific to blocked individuals (compare hashed bars in Figure \ref{fig:figure_quit_social_features_comparison}).  Departing users in our matched pairs do not have significantly different activity levels or activity spreads than their staying counterparts.
This finding reinforces the notion that departure during a block is a distinct phenomenon from departure in general.

\begin{figure*}
    \centering
    \begin{subfigure}[t]{0.225\textwidth}
        \includegraphics[width=\textwidth]{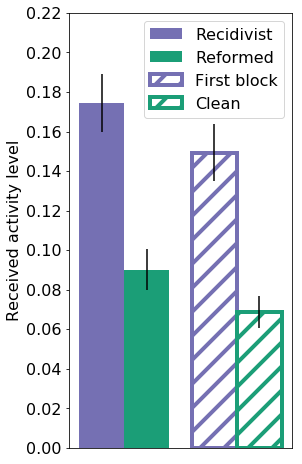}
    \end{subfigure}
    \begin{subfigure}[t]{0.225\textwidth}
        \includegraphics[width=\textwidth]{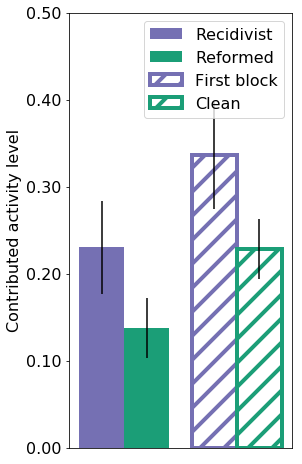}
    \end{subfigure}
    \begin{subfigure}[t]{0.225\textwidth}
        \includegraphics[width=\textwidth]{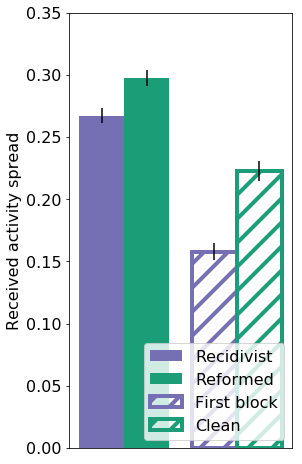}
    \end{subfigure}
    \begin{subfigure}[t]{0.225\textwidth}
        \includegraphics[width=\textwidth]{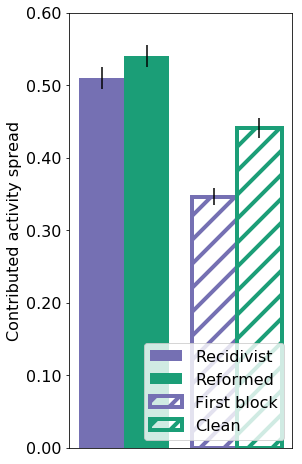}
    \end{subfigure}
    \caption{Difference between the community engagement patterns of recidivist and reformed users (solid bars). From left to right: receiving activity level,  contributed activity level, receiving activity spread, contributed activity spread. For reference, hashed bars compare users that get blocked for the first time with those that never get blocked.}
    \label{fig:figure_recidivism_social_features_comparison}
\end{figure*}

\xhdr{Recidivism}
Having shown that activity level and activity spread are indicative of departure, we now explore whether they might also signal recidivism.
As before, we use matching to eliminate trivial confounds.
In this setting, the primary concern is that higher activity might trivially result in higher likelihood of recidivism merely because users with more activity tend to write more comments, which means they have more things they could be blocked for, probabilistically speaking.
We therefore match each recidivist user by counting the number of actions (new comments, comment edits, or comment deletions) they performed between consecutive blocks, and choosing a reformed user who made at least that many actions in their lifetime after their first (and only) block; this is visualized in Figure \ref{fig:pairing_recidivism} (top).
This ensures that both users had at least as many opportunities to be blocked, so any differences that remain must result from the quality, not quantity, of their actions. This pairing results in 1,332 paired short-term recidivists and reformed users and 3,278 paired long-term recidivists and reformed users.

As in the departure analysis, we also find a set of control pairs to help us understand whether any differences we observe are specific to recidivism, or apply generally to norm violations.
Each control pair contains a first-time offender and a user who is completely clean up to the time of the first-time offender's block.
Selection is done on number of actions, just like the pairing of the experimental group.
This procedure is visualized in Figure \ref{fig:pairing_recidivism} (bottom).

Figure \ref{fig:figure_recidivism_social_features_comparison} summarizes how our engagement factors relate to recidivism and to blocking in general.
For space considerations, we only present results for long-term recidivism;  short term recidivism shows the same trends and statistical significance levels.
Recidivist users receive and contribute more daily comments, but the received activity comes from a narrower set of users (solid bars).
Taken together, these indicate that recidivist users' interaction patterns have more depth than breadth; they engage with a smaller set of users but each engagement is more prolonged.
This may signal the presence of sustained back-and-forth with a restricted circle of editors.  
These engagement patterns echo general blocking patterns (hashed bars): unsurprisingly, the type of individuals that are likely to get blocked in general are also more likely to recidivate (although this was not the case for departure). 
In the following sections we will complement this general observation by exploring factors that relate to the context of the first block, and thus are specific to recidivism.

\xhdr{Predicting future trajectory}
To estimate the relative predictive power of these user characteristics factors and compare them with natural baselines we formulate three new \emph{future} prediction tasks: one for departure, one for long-term recidivism, and one for short-term recidivism.
Each task is run on the respective paired sets of users previously described, resulting in a balanced prediction task.

For transparency, we train a simple SVM classifier with our activity features as input, employing both the normalized and unnormalized versions of the activity level features.
In each prediction task, hyperparameters for the SVM are selected through iteration over different combinations on a held-out development set, and accuracy is reported using leave-one-out cross-validation on the rest of the data.

We consider as features the user characteristics introduced earlier, calculated based only on information available \emph{before} the block: community age and engagement patterns (activity level and spread). We compare their predictive power to two natural baselines that indicate the type and duration of the initial block.  These account for the fact that certain kinds of offenses might be inherently more or less likely to lead to departure or recidivism.

\begin{table}
    \centering
    \begin{tabular}{llll}
        \hline
        & & Long-term\hspace{-3pt} & Short-term  \\
         Feature set & Departure\hspace{-3pt} & recidivism\hspace{-3pt} & recidivism  \\
        \hline
        Baseline: reason & $59.0\%$ & $52.9\%$ & $51.9\%$ \\
        Baseline: duration\hspace{-1.1pt} & $56.7\%$ & $50.0\%$ & $43.8\%$ \\
        \sectionrule
        Community age & $58.6\%$ &$58.7\%^{*}$ & $56.3\%^{*}$ \\
        Engagement patterns\hspace{-8pt} & $61.4\%^{*}$ & $59.0\%^{*}$ & $59.1\%^{*}$ \\
        \sectionrule
        Engagement + Age & $66.2\%^{*}$ & $60.6\%^{*}$ & $58.8\%^{*}$ \\
        \hline
    \end{tabular}
    \caption{Accuracies on three post-block trajectory prediction tasks (random baseline is 50\%). $^*$s indicate results that are significantly outperforming the best performing baseline for the respective task ($p<0.05$).}
    \label{tab:table_prediction_results}
\end{table}

Table \ref{tab:table_prediction_results} summarizes the results of the prediction experiments.
We find that both community age and engagement patterns outperform baselines in all three prediction settings, showing they encode additional information about the future trajectory of the individual.  Furthermore, they also combine to outperform each individual feature, showing that they capture different aspects of the phenomenon. 

Overall, these results demonstrate that forecasting the future post-block trajectory of a user is at least a feasible and well-formed task.  While our main goal in formulating these prediction tasks is to better understand the relative importance of different factors, we advise against using the resulting classifiers for any applications of algorithmically-assisted moderation.  Given the sensitive nature of such applications, a public-facing model would need to be thoroughly tuned to ensure fair outcomes and exhaustively studied to understand if any biases in the data are being encoded \cite{zafar_fairness_2017,feldman_certifying_2015,corbett-davies_algorithmic_2017}.    Understanding the sources of such potential biases in the data, and devising mechanisms to eliminate them, constitutes an important avenue for future work.

\section{Block context}
\label{sec:context}

Having established the relation between general user characteristics and post-block trajectories, we now seek to understand how the nature and context of the block itself mediates recidivism.
Deterrence and defiance theories contrast two possible factors that relate to future compliance~\cite{sherman_defiance_1993}.
The traditional perspective is that the more severe the punishment is, the more likely it is to deter the perpetrator from misbehaving in the future \cite{grasmick_deterrent_1980, klepper_tax_1989}.
More recent work, however, suggests that a factor that is at least as important is whether the perpetrator perceives the punishment to be fair and justified \cite{paternoster_fair_1997,williams_arrest_2005}: intuitively, individuals are unlikely to follow rules that they perceive to be imposed by an unfair and antagonistic society. 
We explore both these perspectives in the context of online recidivism, and bring supporting evidence for the 
perceived fairness 
perspective.

\subsection{Block severity}
The severity of temporary blocks corresponds naturally to their duration.
For the purposes of this analysis, we only consider blocks that never had their duration changed.
Since block durations are highly concentrated on the shorter end of the scale, with extreme outliers being sparsely spread in a long tail, we binarize durations as short (1 day or less) and long (more than 1 day), where the cutoff of 1 day was chosen because it is the median duration.
Perhaps surprisingly, we find no statistically significant difference
(chi-squared test)
between these two classes in terms of recidivism rates, neither for long-term blocks, nor for short-term blocks.

While block duration does not appear to impact future compliance, it does correlate with departure rates: 30\% of the users receiving long blocks quit within one month after the start of the block, compared with 24\% for short blocks (significant at $p < 0.001$).\footnote{Rather than analyzing departure during of the block as in the previous analysis, we are constrained here to consider a fixed time window in order to account for the time a user has to depart.} This result holds even when controlling for the age and activity of the blocked individual.%

\begin{table*}[ht]
    \centering
    \begin{tabular}{ll}
        \hline
        Feature & Examples \\
        \hline
        Apologizing & It is true that the tag-war got out of hand. I \textbf{apologize} if it caused any disruption.\\
        & I am deeply \textbf{sorry} for not understanding the whole situation, and ask for your forgiveness. \\
        & Given the recent history on the article, \textbf{forgive me} if claims of lawyers are met with skepticism. \\
        \sectionrule
        Direct questioning & \textbf{Why} was I blocked, when I had not the right, but rather the duty to remove BLP inaccuracies?\\
        & \textbf{So what} policy, precisely have I violated? \\
        & \textbf{How} is it this person won't leave me alone and I repeatedly ask to be left alone and I get blocked?\\
        \sectionrule
        Unfairness lexicon & this block is \textbf{unjustified}. none of the changes were in violation of the 3rr rule. \\
        & i have alerted another administrator about your blatent and \textbf{unwarranted} abuse of power. \\
        & iv been \textbf{wrongly accused} of making up information \\
        \hline
    \end{tabular}
    \caption{Example blocked user messages exhibiting linguistic indicators of (un)fairness perception.  Bolded words indicate matched patterns.}
    \label{tab:table_feature_examples}
\end{table*}

\begin{figure*}
    \centering
    \begin{subfigure}[t]{.225\textwidth}
        \includegraphics[height=1.6in]{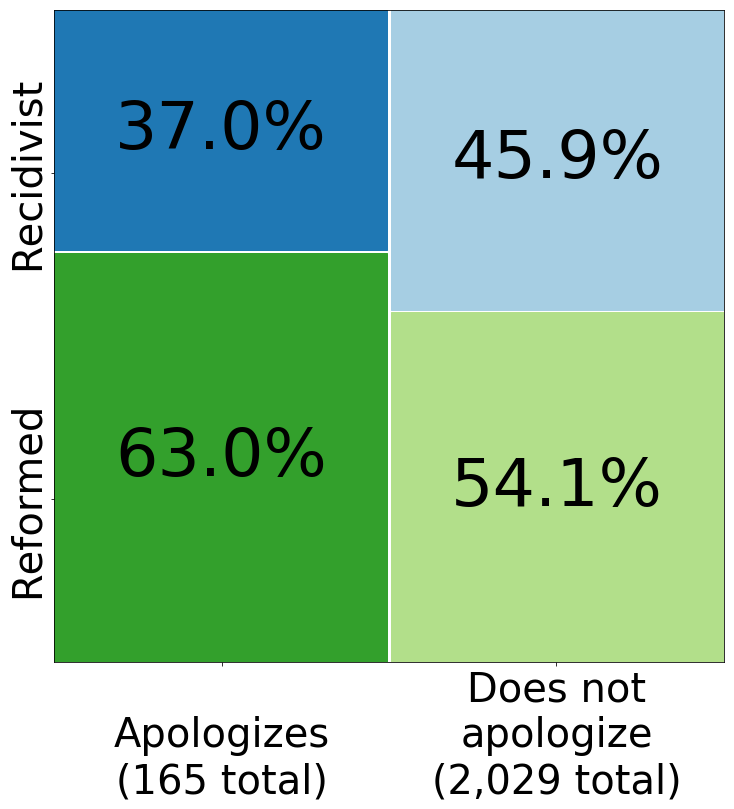}
        \subcaption{Apologizing}
        \label{fig:figure_mosaics_apologizing}
    \end{subfigure}
    \begin{subfigure}[t]{.225\textwidth}
        \includegraphics[height=1.6in]{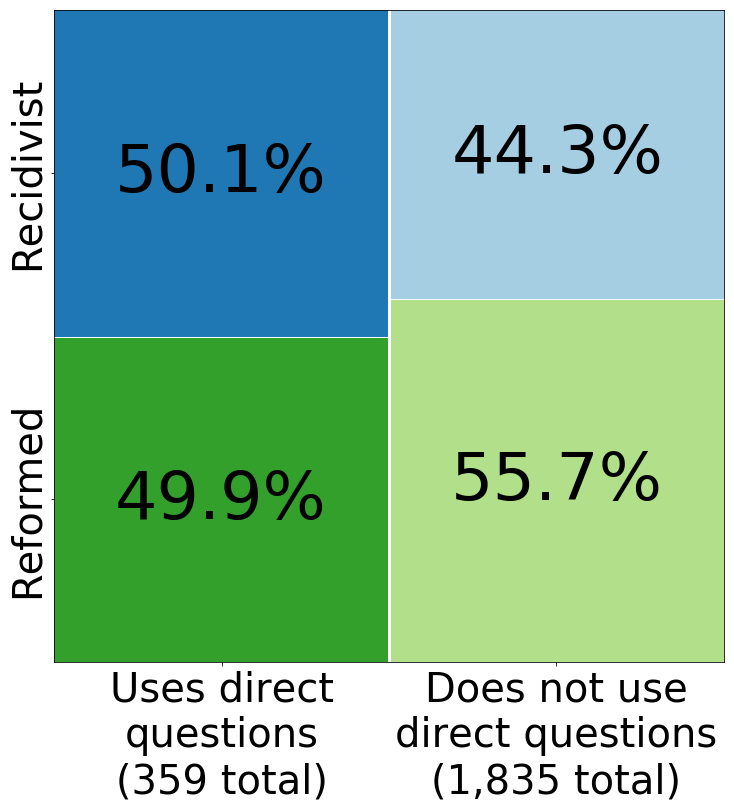}
        \subcaption{Direct questioning}
        \label{fig:figure_mosaics_direct_questions}
    \end{subfigure}
    \begin{subfigure}[t]{.225\textwidth}
        \includegraphics[height=1.6in]{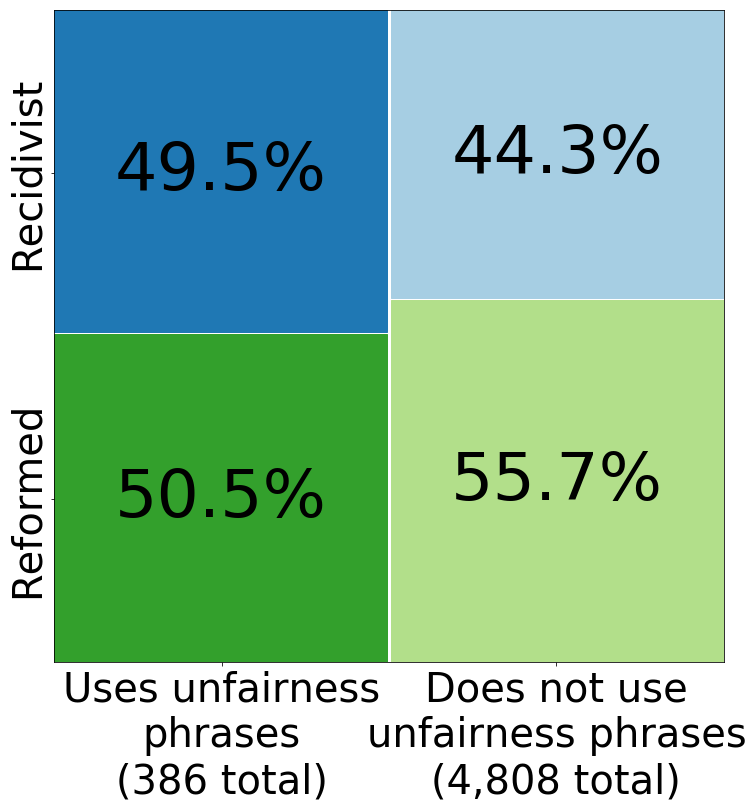}
        \subcaption{Unfairness lexicon}
        \label{fig:figure_mosaics_unfairness}
    \end{subfigure}
    \begin{subfigure}[t]{.225\textwidth}
        \includegraphics[height=1.6in]{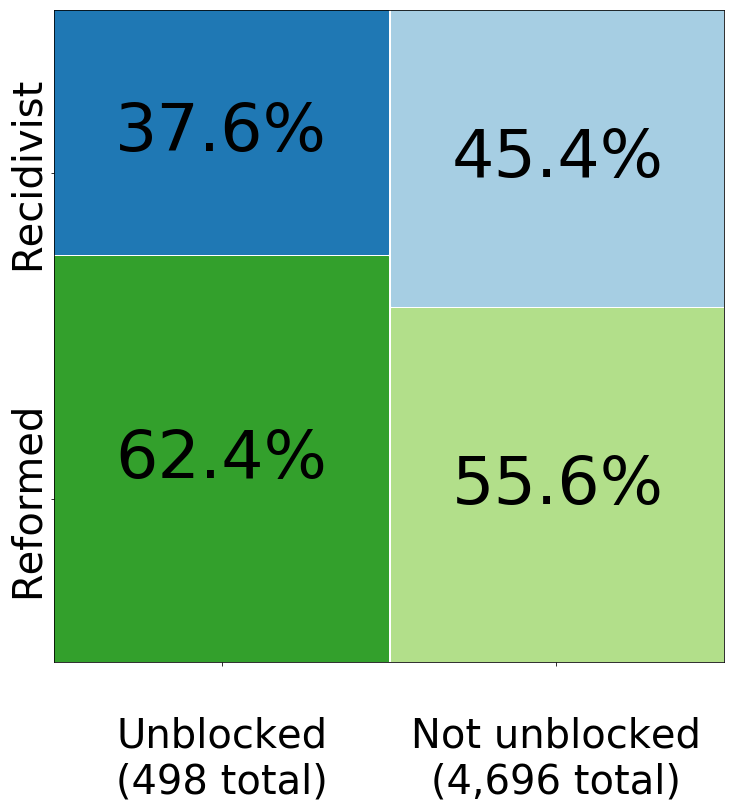}
        \subcaption{Unblocks}
        \label{fig:figure_mosaics_unblock}
    \end{subfigure}
    \caption{Normalized contingency tables showing the relative ratios of recidivist and reformed users conditioned on the presence or absence of fairness perception indicators. Table cell heights are scaled to the frequencies they represent in order to facilitate comparison between the two conditions (compare left to right in each figure).}
    \label{fig:figure_mosaics}
\end{figure*}

\subsection{Perceived fairness}
Considering that in our online setting the actual severity of a block does not appear to impact its effectiveness, we can investigate how the way a user \emph{perceives} their block relates to their future compliance with community rules.\footnote{While for brevity we only discuss long-term recidivism, the same effects and statistically significant levels hold for short-term recidivism.}

The main challenge in addressing questions related to user perception is capturing such a subjective notion at a sufficiently large scale.
Unlike offline studies that rely on interviews with punished individuals to assess their perception of the fairness of the punishment \cite{paternoster_fair_1997}, our observational data does not provide direct access to such variables.
We can, however, leverage a specific Wikipedia mechanism in order to estimate them for a subset of users: block appeals.

A block appeal is a formal avenue through which a blocked user can contest the block, admit their wrongdoing, or ask for clemency. 
In response to an appeal, any moderator---not necessarily the one responsible for the original block---may choose to reduce the duration of the block. 
This happens rather rarely, with only 15\% of blocked users having their block duration reduced.  As described on Wikipedia's official policy page on block appeals,\footnote{\url{https://en.wikipedia.org/wiki/Wikipedia:Appealing_a_block}} there are two reasons a moderator may choose to grant an appeal. The most common reason is clemency: if the blocked user takes responsibility for their offense, acknowledges that they were in the wrong, and convincingly argues that they will not offend again, the reviewing moderator may grant a shortened block.  The other reason is if the moderator judges the original block to be in error, e.g., due to a case of mistaken identity.
However, according to the cited Wikipedia policy, the latter scenario is ``extremely rare'', due in part to the public nature of the moderation in Wikipedia and to the strict procedure which blocking moderators need to undergo in order to block somebody, which involves identifying a specific reason.

\xhdr{Linguistic signals} Block appeals provide two separate opportunities to (admittedly imperfectly) estimate how users perceive the block.  The first one is through the language the blocked individual uses in their appeals.  They can signal whether they perceived the block as being justifiable and acknowledge the wrongdoing or, conversely, whether they disagree with the justification for the block and contest it. 

To capture such signals we consider three types of linguistic cues (Table \ref{tab:table_feature_examples}) that are intuitively related to the user's perception of the fairness of the block.
Users that apologize signal that they acknowledge their wrongdoing; we expect such users to perceive the block as being fair and thus to be less likely to recidivate. 
Conversely, users that openly and directly challenge their block, either by aggressively questioning the administrators or by explicitly claiming unjust treatment, signal that they perceive the block as unfair and thus we would expect that they are more likely to recidivate.

To capture apologizing and direct questioning we use off the shelf tools\footnote{From ConvoKit: \href{http://convokit.cornell.edu}{convokit.cornell.edu}} that perform regular expression matching on dependency trees  
\cite{danescu-niculescu-mizil_computational_2013}.
To capture explicit claims of unfairness we experimented with hand-building a small lexicon by examining a sample of examples from the data:\footnote{We also performed a Fightin' Words analysis \cite{monroe_fightin_2008} to identify salient cues not present in the sample data we examined.} ``unjust'', ``illegitimate'', ``illegal'', ``unfair'', ``not fair'', ``accuse'' (and different forms), ``wrongly'', ``falsely'', ``injustice'', ``unfounded'', ``allege'' (and different forms), and ``unwarranted''.

We extract these linguistic cues from the messages recidivist and reformed users direct at moderators during their block, capturing their use in both formal and informal appeals.  Figure \ref{fig:figure_mosaics} (a-c) compares the relative ratios of recidivist and reformed users among those that used a given type of fairness cue or not.  Compared to a baseline ratio of 45\% recidivism and 55\% reform that consistently holds in all three cases, apologizing corresponds to a reduction in the rate of recidivism (chi-squared test $p < 0.05$), while direct questioning   and explicitly calling for unfairness correspond to a moderate increase in the rate of recidivism ($p < 0.05$ and $p = 0.053$, respectively).

\xhdr{Granted appeals} A second opportunity to estimate the user's perception of a block is by analyzing \emph{granted} appeals. When an appeal is granted, the duration of the block is shortened.  This positive response from the moderators arguably signals to the user that the system is fair and that their views are taken seriously. \citeauthor{paternoster_fair_1997} (\cite{paternoster_fair_1997}, p. 169) shows that such perceptions about the enforcing authorities decrease the rate of recidivism offline, thus we hypothesize that this could hold true in  online settings as well.  Figure~\ref{fig:figure_mosaics_unblock} shows that this principle is indeed echoed on Wikipedia; users who are unblocked are significantly more likely to reform $(p < 0.001)$.   This effect is even more striking when considering that, as discussed above, the duration of the block itself has no noticeable effect on recidivism and 
that granted appeals
often shorten the duration of the block by less than one day (the median reduction is 22 hours).

\xhdr{Alternative explanations} One explanation that does not relate to perceived fairness could be that these appeal-based effects are simply differentiating between individuals who did not deserve to be blocked in the first place (and thus did not commit an offense and are unlikely to offend at all) and those that did commit an offense.  While our observational approach precludes ruling out this alternative explanation completely, we argue that it is at least unlikely. First, as noted earlier, it is extremely rare for users to be blocked in error on Wikipedia. Second, some of the linguistic effects go in the opposite direction of what such an explanation would predict: for example, apologizing generally comes with an admittance of guilt, and under the alternative explanation we would not expect such guilty individuals to have a reduced rate of recidivism.
In fact, we find that individuals who apologize are significantly ($p < 0.001$, by chi-squared test) more likely to get unblocked than ones who do not, a pattern that is consistent with what we would expect if unblocks tend to be granted because of clemency (acknowledgment of wrongdoing being one of the factors that moderators look for when deciding whether to grant an appeal).

In an attempt to further discard explanations based on user characteristics we also repeat these experiments after controlling for age and activity (i.e., pairing recidivist and reformed users based on these variables before conducting the analysis) and found similar trends, with exception of the unfairness lexicon (which had a moderate effect to start with and was likely sensitive to the sample size reduction).  

We argue that these observations, while not of causal nature, at least motivate further investigations into the role of perceived fairness in online moderation.

\section{Further Related Work}
\label{sec:related}

\xhdr{Antisocial behavior}
Online moderation largely addresses the problem of antisocial behavior, which occurs in the form of harassment \cite{vitak_identifying_2017}, cyberbullying \cite{singh_they_2017}, and general aggression \cite{kayany_contexts_1998}.
Approaches to moderating such content include decentralized, community-driven methods \cite{geiger_bot-based_2016}, as well as top-down methods relying on designated community managers or moderators \cite{mcgillicuddy_controlling_2016}.
Prior research in this area ranges from understanding the actors involved in antisocial behavior \cite{volkova_identifying_2017, cheng_antisocial_2017, ribeiro_characterizing_2018,munger_tweetment_2017,kumar_community_2018} to analyzing its effects \cite{olteanu_effect_2018} to 
 tools for identifying such behavior \cite{wulczyn_ex_2017, nobata_abusive_2016}, and even forecasting future instances 
\cite{zhang_conversations_2018, liu_forecasting_2018}. 
While inspired by this line of work, our present study extends it by focusing on what happens \emph{after} a user is blocked for violating community rules.

\xhdr{Effects of moderation}
Prior work has examined the effects of different kinds of moderation on online platforms.
Community-driven moderation can affect overall user participation \cite{muchnik_social_2013, cunha_effect_2016} and comment quality \cite{cheng_how_2014} in conversations.
Centralized moderation can also have effects on conversation, mainly through the moderator's role as an authority figure \cite{seering_shaping_2017}.
These existing studies of moderation effects have focused on the short term, largely operating at the level of individual conversations.
By contrast, in this work we intend to study the long term, \emph{user} level effects of moderation:
what happens to a user in the days, weeks, and months after a moderator takes action against them?

\xhdr{Norms and engagement}
A major factor governing engagement in online communities is a sense of belonging \cite{newell_user_2016}, which in many communities engenders the emergence of community-specific norms, such as specific patterns of language 
\cite{danescu-niculescu-mizil_no_2013, chancellor_norms_2018,hamilton_loyalty_2017,zhang_community_2017}.
Wikipedia is no exception: it relies on group dynamics to promote editing productivity \cite{kittur_herding_2009} and quality \cite{pavalanathan_mind_2018}, as well as participation in governance \cite{burke_mopping_2008,leskovec_governance_2010}.
\xhdr{Fairness in moderation}
Moderation is fundamentally a difficult process, sparking discussions about design and best practices \cite{kiesler_regulating_2012, castillo_current_2017}, and moral questions regarding bias \cite{shen_perceptions_2018} and free speech \cite{jhaver_view_2018}.
One 
interesting
 avenue of research 
in this broad space
  deals with understanding the \emph{perception} of fairness.
Prior work has 
looked at this from the
 perspective of moderators, showing that 
  moderators on the technology news forum Slashdot 
 frequently judge each others' actions as unfair, but that this rarely leads to reversal of moderation decisions \cite{lampe_slashdot_2004}.
We instead explore the perspective of users, exploring ways to determine when a user regards a moderation decision as unfair.

\section{Discussion}
\label{sec:discussion}
We introduced a computational framework for characterizing the possible trajectory a user can take after a temporary block: departure, recidivism, or redemption.
Taking inspiration from theories of offline punishment,
our investigation focused on two types of factors that are tied to these outcomes: those related to a user's involvement in the community, and those related to their perception of the moderation action affecting them.\footnote{We 
refrain from drawing any explicit connections between online and offline punishment, acknowledging their different nature and goals.} 
 Overall, our work builds towards a more nuanced approach to moderation that 
more broadly accounts for the tradeoffs between possible outcomes and that considers how the affected individuals might perceive the moderators and their actions.

Our study has several limitations that naturally translate into opportunities for  future work.
First, while we have focused on the blocked users, it would be valuable to also understand the normative effect of blocks on other users that are witnessing them.   Additionally, our study is limited to revealing correlational effects, and different methodologies (involving surveys or experimentation) would be necessary to study the underlying causal mechanism.%

In terms of methodology, there are several ways in which the definitions of the post-block trajectories themselves could be refined. While we measure departure as a binary value, in reality community participation occurs on a variable scale and it is possible that blocked users do not depart, but drastically reduce their participation 
 (or conversely, increase it).
Similarly, defining recidivism in terms of a second block is an imperfect approach because it is possible that some users reoffend without getting blocked.
This is especially true in the case of personal attacks, where each individual's threshold of acceptable behavior may vary.
One possible way of addressing this would be to combine our approach with prior work in automated detection of toxic language \cite{davidson_automated_2017,wulczyn_ex_2017,ribeiro_characterizing_2018}.

We also note possible ways in which the factors we explore could be expanded upon.
The activity features we introduced in Section \ref{sec:activity} effectively serve as crude approximations of network effects.
Future work could apply a more formal network analysis to uncover more powerful representations of activity.
Similarly, our linguistic measures of fairness are currently limited.
One immediate way of expanding them would be to build a more comprehensive lexicon of unfairness, perhaps using a semi-automated approach based on word co-occurrences with the currently existing lexicon items.
A more sophisticated approach might involve engineering dependency-based features similar those used to extract apologies and direct questioning. Understanding whether a user feels they are treated unfairly constitutes an interesting research question on its own.

{\small
\xhdr{Acknowledgements}
The authors thank Lucas Dixon, Liye Fu, Jon Kleinberg, Lillian Lee, Dario Taraborelli, Justine Zhang and Andrew Wang and the anonymous reviewers for their helpful comments. This work is supported in part by a Google Faculty Research Award, 
NSF CAREER award IIS-1750615 and
 NSF Grant SES-1741441.
}

\bibliographystyle{ACM-Reference-Format}
\bibliography{recidivism-www-autoupdate,recidivism-www-autoupdate-jpc}
\end{document}